\documentclass[11pt]{amsart}
\usepackage{geometry}                
\usepackage{graphicx}
\usepackage{amssymb}
\usepackage{epstopdf}
\usepackage{array}
\usepackage{subfigure}

\usepackage{supertabular}
\usepackage{hhline}
\newcommand\arraybslash{\let\\\@arraycr}
\makeatother
\newtheorem{finding}{Finding}

\usepackage{lscape}

\title[Excitable automata with dynamical excitation intervals]{On diversity of configurations generated  by excitable cellular automata\\ with dynamical excitation intervals}
\author[Adamatzky]{Andrew Adamatzky}
\address[Adamatzky]{University of the West of England, Bristol, UK}
\email{andrew.adamatzky@uwe.ac.uk}

\begin{document}

\maketitle

\begin{abstract}

Excitable cellular automata with dynamical excitation interval exhibit a wide range of space-time dynamics based 
on an interplay between  propagating excitation patterns which modify excitability of the automaton cells. Such 
interactions leads to formation of standing domains of excitation, stationary waves and localised excitations. We analysed 
morphological and generative diversities of the functions studied and characterised the functions with highest values of 
the diversities. Amongst other intriguing discoveries we found that upper boundary of excitation interval more significantly affects morphological diversity of  configurations generated than lower boundary of the interval does and there is no match between functions which produce configurations of excitation with highest morphological diversity and configurations of interval boundaries with highest morphological diversity.  Potential directions of future studies of excitable media with dynamically changing excitability may focus  on relations of the automaton model with living excitable media, e.g. neural tissue and muscles, novel materials with memristive properties, and networks of 
conductive polymers.

\end{abstract}

\section{Introduction}

Since their popularisation in \cite{greenberg-hasting}, excitable cellular automata 
became a convenient tool for studying complex phenomena of excitation dynamics and 
chemical reaction-diffusion activities in physical, chemical and biological systems~\cite{ilachinski,chopard}. 
The cellular automata offers quick 'prototyping' of complex spatially extended non-linear media. 
The examples of `best practice' include models of Belousov-Zhabotinsky reactions and other excitable
systems~\cite{gerhardt,markus_hess_1990}, chemical systems exhibiting Turing patterns~\cite{young,yaguma,yang2004},  precipitating systems~\cite{adamatzky_book_2005}, calcium wave 
dynamics~\cite{yang}, and chemical turbulence~\cite{hartman}.

In a classical Greenberg-Hasting~\cite{greenberg-hasting} automaton model of excitation a cell takes 
three states --- reseting, excited and refractory. A resting cell becomes excited if number of excited neighbours
exceeds a certain threshold, an  excited cell becomes refractory, and a refractory cell returns to its original 
resting state.  In 1998~\cite{adamatzky_holland_1998}, we introduced an excitable cellular automaton, where a resting
cell is excited if a number of its excited neighbours belongs to some fixed interval $[\theta_1, \theta_2]$. The 
interval $[\theta_1, \theta_2]$ was called an excitation interval. For a two-dimensional cellular automaton with 
eight-cell neighbourhood boundaries of the excitation interval satisfy the condition: 
$1 \leq \theta_1 \leq \theta_2 \leq 8$.  We found that by tuning $\theta_1$ and $\theta_2$ we can force the automaton to imitate almost all kinds of excitation dynamics, from classical target and spiral waves observed in physical and chemical excitable media to wave-fragments inhabiting sub-excitable media. 

How does excitation dynamics change if we allow boundaries of the excitation interval to change during the automaton development? We partially answer the question in present paper by making the interval $[\theta_1^t(x), \theta_2^t(x)]$ of every cell $x$ to be dynamically updatable at every step $t$ depending on state of the cell $x$ and numbers of excited and refractory neighbours in the cell $x$'s neighbourhood. 

The excitable automata with dynamical excitation intervals are defined in Sect.~\ref{definition}. Morphological diversity of cellular automata (measured using Shannon entropy and Simpson index) with different functions of interval updates is characterised in Sect.~\ref{morphologicaldiversity}. Section~\ref{Generativediversityandlocalisations}  characterises generative diversity (measured in terms of different configurations generated during space-time development of
automaton starting with a single non-resting cell) of the local transitions. Some afterthoughts are offered 
in Sect.~\ref{summary}.

\section{Dynamical excitation intervals}
\label{definition}

Let $x^t$ and $x^{t+1}$ be states of a cell $x$ at time steps $t$ and $t+1$, and  $\sigma^t_+(x)$ be a sum of 
excited neighbours in cell $x$'s neighbourhood $u(x)=\{ y : |x - y |_{L_\infty} = 1 \}$.  Cell $x$ updates its state by the following rule:
$$
x^{t+1}=
\begin{cases}
+, \text{ if } x^t=\cdot \text{ and }  \sigma_+^t(x)+ \in [\theta_1^t(x), \theta_2^t(x)] \\
-, \text{ if } x^t=+ \\
\cdot, \text{ otherwise } 
\end{cases}
$$
A resting cell is excited if number of its neighbours belongs to excitation interval $[\theta_1^t(x), \theta_2^t(x)]$,  
where $1 \leq \theta_1^t(x), \theta_2^t(x) \leq 8$.  The boundaries $\theta_1^t(x)$  and $\theta_2^t(x)$ are 
dynamically updated depending on cell $x$'s state and numbers of $x$'s excited $\sigma^t_+(x)$ and 
refractory $\sigma^t_-(x)$ neighbours. A natural way to update boundaries is by increasing or decreasing
their values as follows: 
$$
\theta_1^{t+1}(x) = \xi(\theta_1^t(x) + \Delta_1 \phi(\sigma^t_+(x) - \sigma^t_-(x))) 
$$
$$
\theta_2^{t+1}(x) = \xi(\theta_2^t(x) + \Delta_2 \phi(\sigma^t_+(x) - \sigma^t_-(x))) 
$$
where
$$
\begin{array}{cc}
\Delta_1 = 
\begin{cases}
T_1, \text{ if } x=+ \\
T_3, \text{ if } x=- \\
0, \text{ if } x=0
\end{cases}
&
\Delta_2 = 
\begin{cases}
T_2, \text{ if } x=+ \\
T_4, \text{ if } x=- \\
0, \text{ if } x=0
\end{cases}
\end{array}
$$
and $\phi(a - b)=1$ if $a>b$, 0 if $a=b$ and -1 if $a<b$, and  $\xi(a)=1$ if $a<1$ and 8 if $a>8$.  
Boundaries of excitation interval $[\theta_1^t(x), \theta_2^t(x)]$ are updated independently of each other. 
Local excitation rules are determined by values of $T_1, \cdots T_4$. We therefore address the functions as tuples $E(T_1 T_2 T_3 T_4)$ which range from $E(-1-1-1-1)$ to $E(1111)$. 

Functions $E(a0b0)$, $a \in \{-1, 0, 1 \}$ represent rules with  fixed upper boundary $\theta_2$ of excitation and 
dynamically updated lower boundary $\theta_1$. These are equivalent to dynamically updated thresholds of excitation. 
Functions $E(0a0a)$, $a \in \{-1, 0, 1 \}$ represent rules with fixed lower boundary and dynamical upper boundary of 
excitation interval. 

The experiments are conducted on a cellular array of $n \times n$ cells with periodic boundary conditions. 
In a typical experiment we perturb resting cellular array with a localised domain of excitation, wait till transient 
period is over (1000 iterations  is enough) and most excitation patterns collide and disappear and persist indefinitely, 
and then analyse three configurations: configuration of excitation represented by an array of cells states $x^t$, and configurations
of lower $\theta^t_1(x)$ and upper $\theta^t_2(x)$ boundaries of excitation intervals.

Initially $\theta^0_2(x)=8$ for any $x$. In experiments we considered initial conditions $\theta^0_1(x)=1$ and $\theta^0_1(x)=2$. 
The following scenaria of initial excitation were played:
\begin{itemize}
\item ($++$)-start, $\theta_1=2$: all cells are resting but two neighbouring cells are assigned excited state,
\item R1-start: let $D$ be a disc radius $n/4$ centred in the array $L$ of $n \times n$ cells, all cells are resting but 
cells lying in $D$ are assigned excited states with probability 0.2 and $\theta^0_1(x)=1$ for any $x$,
\item R2-start: all cells are resting but cells lying in $D$ get excited states with probability 0.2 and $\theta^0_1(x)=2$ for any $x$,
\item D1-start:  all cells are resting but cells lying in $D$ get excited states with probability 0.1 or refractory states with probability 
0.1 and $\theta^0_1(x)=1$ for any $x$,
\item D2-start:  all cells are resting but cells lying in $D$ get excited states with probability 0.1 or refractory states with probability 0.1 
and $\theta^0_1(x)=2$ for any $x$,
\item ($-+$)-start: all cells are resting but one cell is excited and its western neighbour is refractory,
\item ($-++$)-start all cells are resting but one cell is excited, its first order western neighbour is excited and
its second order western neighbour is refractory. 
\end{itemize}

Cell states  were represented by colours and grey levels as follows: excited state $+$ is red (c. 76 grey), resting state is white and refractory state $-$ is blue (c. 28 grey). Colour values of  excitation interval boundaries $\theta_1$ and $\theta_2$  are following: 1 is white, 2 is green or   150 grey, 3 is yellow or 226 grey, 4 is blue or 28 grey, 
5 is magenta or 104 grey,  6 is cyan or 178 grey, 7 is red or 76 grey, and 8 is black.

\section{Morphological diversity}
\label{morphologicaldiversity}

We evaluated morphological diversity of configurations of excitation and using Shannon entropy and Simpson's index. 
Let $W=\{ \circ, +, - \}$ be a set of all possible configurations of a 9-cell neighbourhood $w(x)=u(x) \cup x$, 
$x \in \mathbf{L}$. Let $c$ be a configuration of automaton, we calculate number of non-resting neighbourhood 
configurations as $\eta = \sum_{ x \in \mathbf{L}} \epsilon(x)$, where $\epsilon(x)=0$ if for every resting $x$ all 
its neighbours are resting, and $\epsilon(x)=1$ otherwise. The Shannon entropy is calculated as 
$\mathcal{E}=- \sum_{w \in W} (\nu(w)/\eta \cdot ln (\nu(w)/\eta))$, where $\nu(w)$ is a number of times the neighbourhood 
configuration $w$ is found in automaton configuration $c$. Simpson's index is calculated as $S=1- \sum_{w \in W} (\nu(w)/\eta)^2$. 
The measures $\mathcal{E}$ and $S$ were calculated on configurations of cell-states and interval boundaries after
long transient period, sufficient enough for any perturbation to settle down.

\begin{figure}[!tbp]
\centering
\includegraphics[width=\textwidth]{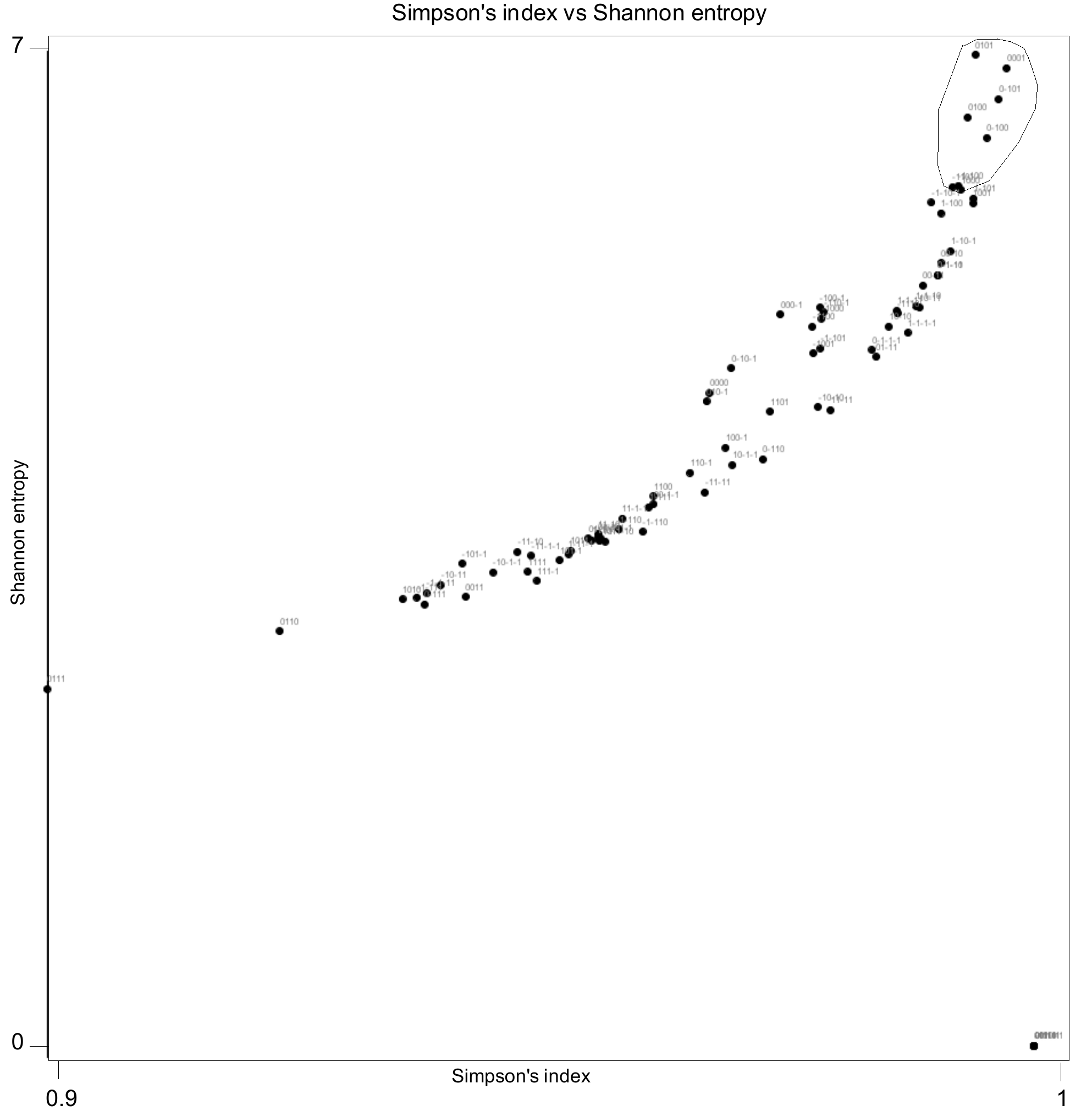}
\caption{Morphological diversity of functions for D1-start, $\theta^0_1(x)=1$ for all $x$: 
Simpson's index $S$ (horizontal axis) vs Shannon entropy $\mathcal{E}$ (vertical axis) for configuration of excitable array of $200 \times 200$ cells with periodic boundary condition, recorded at $t=1000$. 
Encircled data points are seven functions with highest morphological diversity specified in column D1-start, 
$S$-$\mathcal{E}$ in Tab.~\ref{diversitytable}.}
\label{simpson_shanon_randomdisc_T1=1}
\end{figure}

\begin{figure}[!tbp]
\centering
\includegraphics[width=\textwidth]{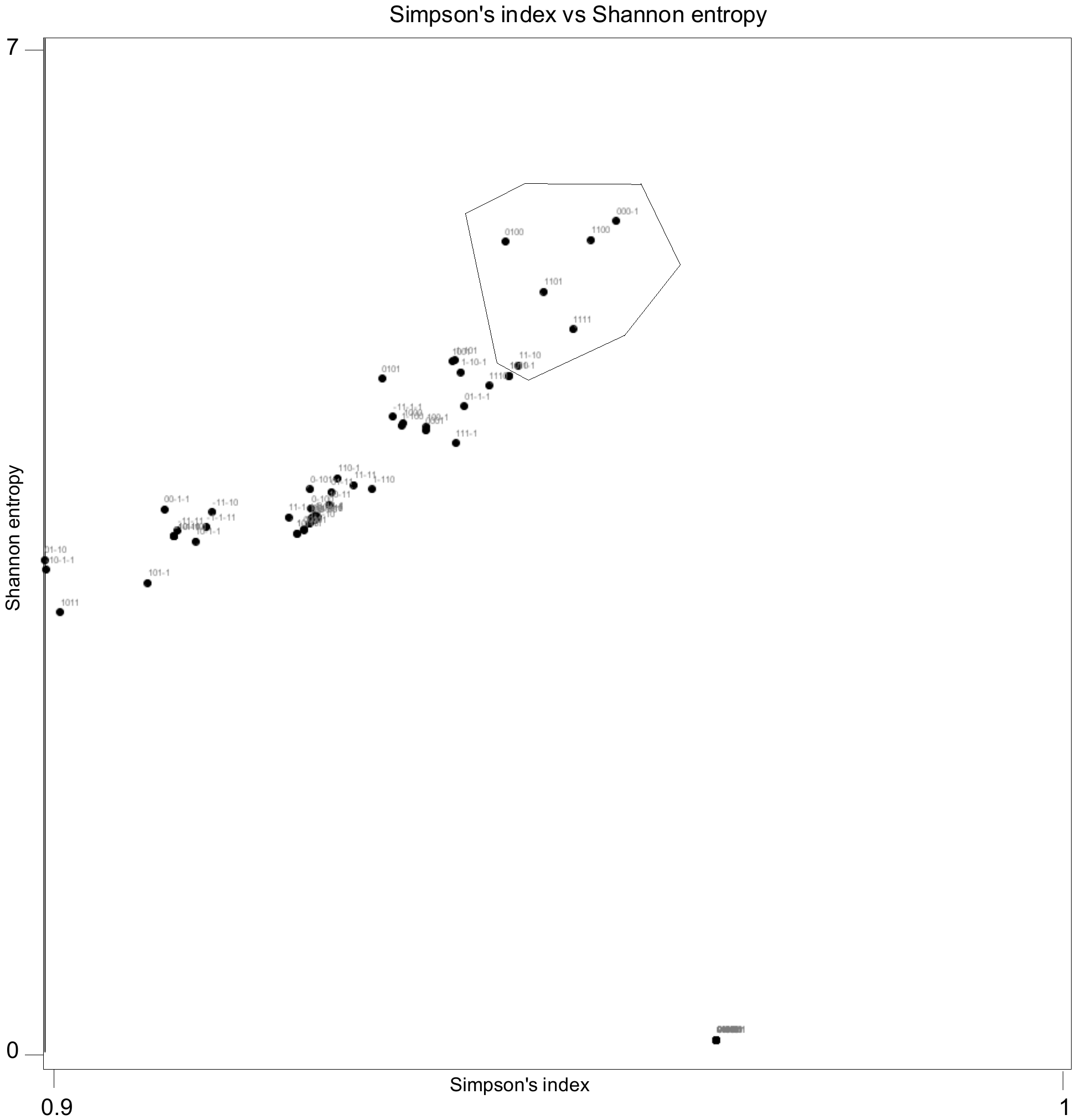}
\caption{Morphological diversity of functions for ($++$)-start, $\theta^0_1(x)=2$ for all $x$: 
Simpson's index $S$ (horizontal axis) vs Shannon entropy $\mathcal{E}$ (vertical axis) for configuration of excitable array 
of $200 \times 200$ cells with periodic boundary condition, recorded at $t=1000$. 
Encircled data points are seven functions with highest morphological diversity specified in column ($++$)-start, 
$S$-$\mathcal{E}$ in Tab.~\ref{diversitytable}.}
\label{simpson_shanon_singleton_T1=2}
\end{figure}

\begin{figure}[!tbp]
\centering
\includegraphics[width=\textwidth]{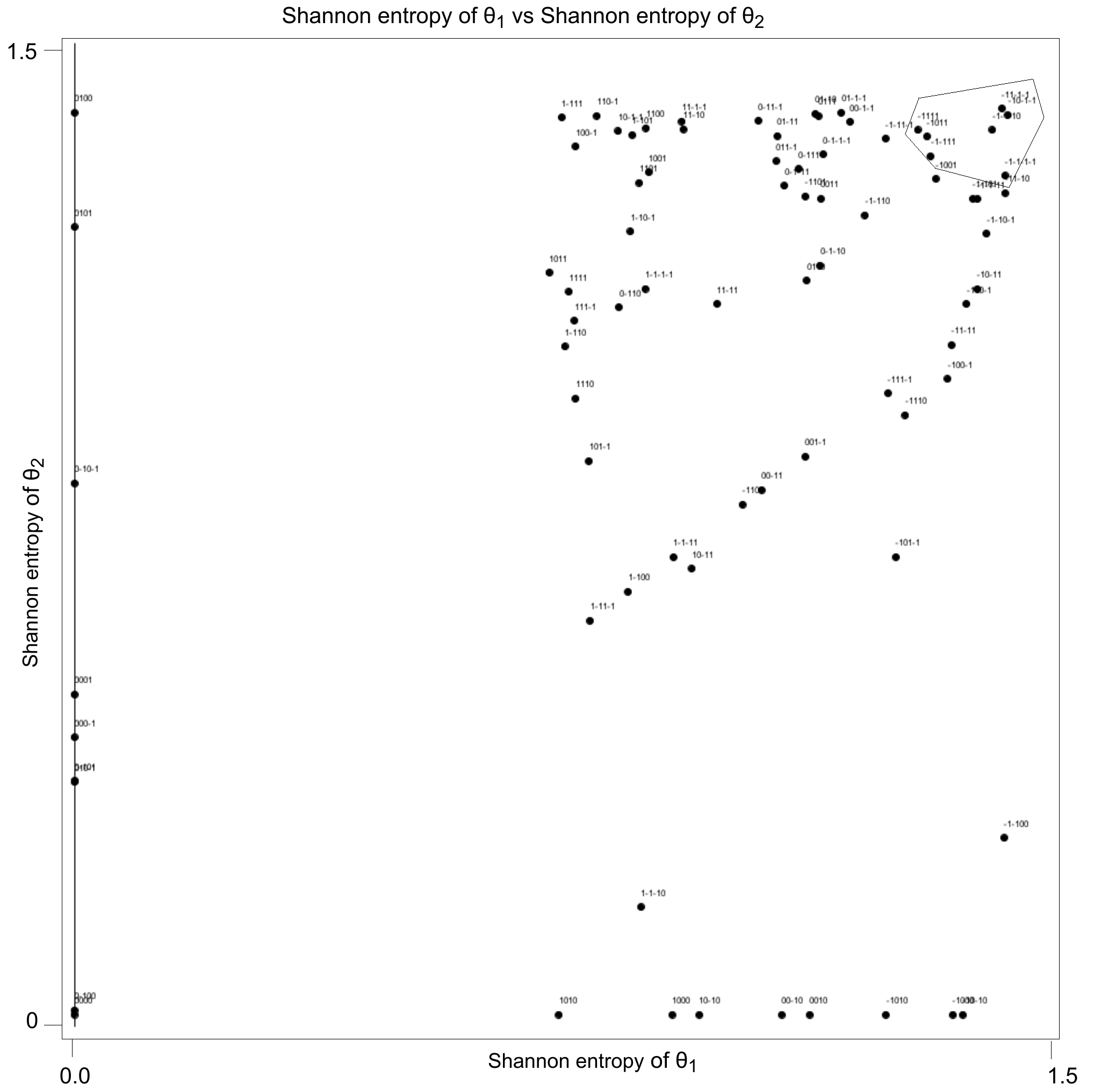}
\caption{Morphological diversity of functions for D1-start, $\theta^0_1(x)=1$ for all $x$: 
Shannon entropy $\mathcal{E}_1$ for configuration of $\theta_1$ (horizontal axis) vs  Shannon entropy 
$\mathcal{E}_2$ for configuration of $\theta_2$ (vertical axis), 
recorded at $t=1000$. 
Encircled data points are seven functions with highest morphological diversity specified in column D1-start, 
$\mathcal{E}_1$-$\mathcal{E}_2$ in Tab.~\ref{diversitytable}.}
\label{shannon1_shanon2_randomdisc_T1=1}
\end{figure}

\begin{figure}[!tbp]
\centering
\includegraphics[width=\textwidth]{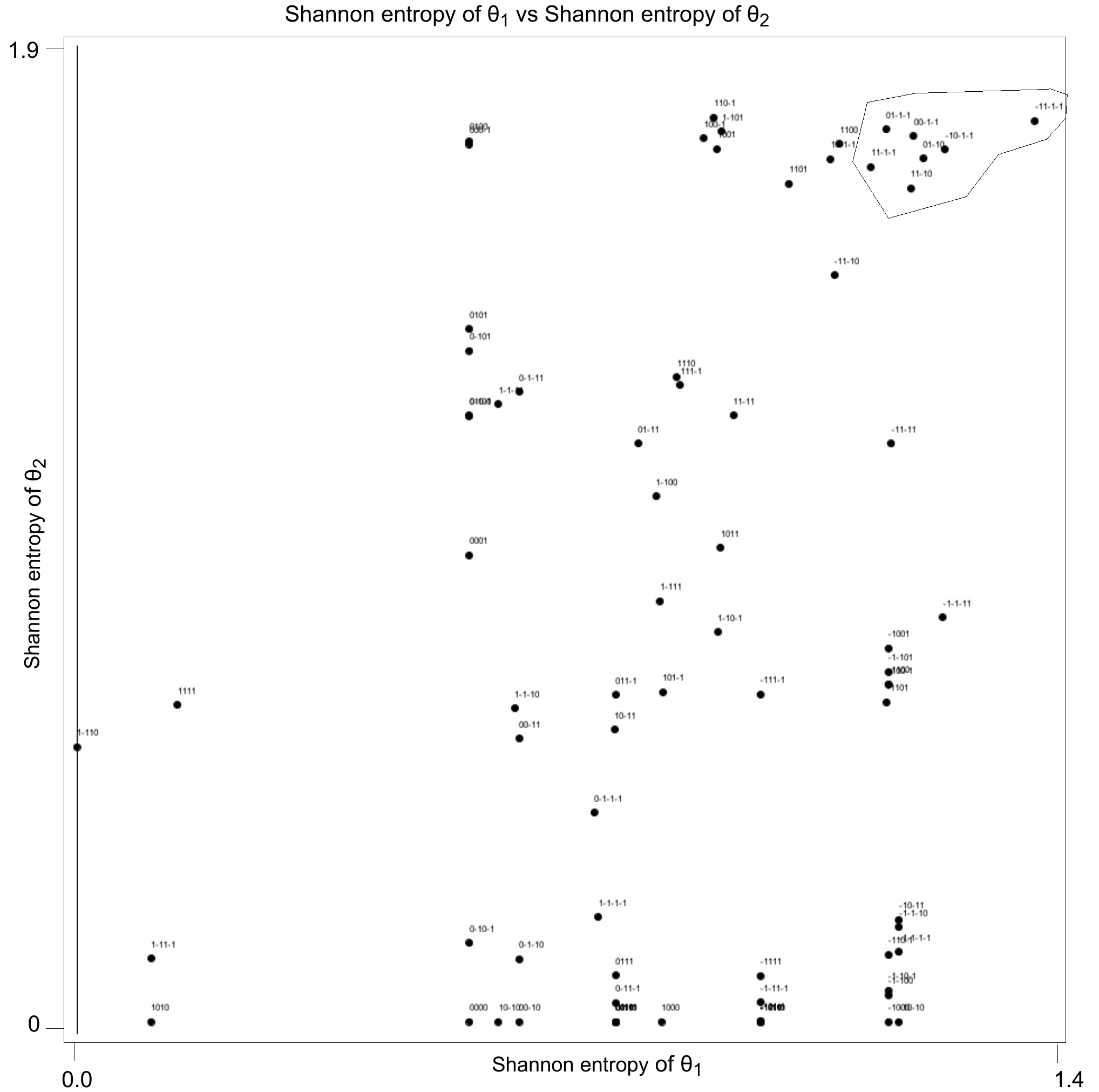}
\caption{Morphological diversity of functions for ($++$)-start, $\theta^0_1(x)=2$ for all $x$: 
Shannon entropy $\mathcal{E}_1$ for configuration of $\theta_1$ (horizontal axis) vs Shannon entropy $\mathcal{E}_2$ for 
configuration of $\theta_2$ (vertical axis),  recorded at $t=1000$.  Encircled data points are seven functions 
with highest morphological diversity specified in column ($++$)-start, $\mathcal{E}_1$-$\mathcal{E}_2$ in Tab.~\ref{diversitytable}.}
\label{shannon1_shanon2_singleton_T1=2}
\end{figure}

The diversity of excitation patters is evaluated using $S$-$\mathcal{E}$ plots. See examples of $S$-$\mathcal{E}$ plots for 
D1-start, $\theta^0_1(x)=1$, in Fig.~\ref{simpson_shanon_randomdisc_T1=1} and ($++$)-start, $\theta^0_1(x)=2$,
in Fig.~\ref{simpson_shanon_singleton_T1=2}.  Distributions of functions by their values of Shannon entropies for 
$\theta_1$ and $\theta_2$ are illustrated in Figs.~\ref{shannon1_shanon2_randomdisc_T1=1} 
and \ref{shannon1_shanon2_singleton_T1=2}



\begin{landscape}
\centering
\begin{table}[!tbp]
\begin{tabular}{|m{0.9cm}|m{1.1cm}|m{0.7cm}|m{1.1cm}|m{1cm}|m{1.2cm}|m{1cm}|m{1.2cm}|m{1cm}|m{1.1cm}|m{1.1cm}|m{1.2cm}|m{1cm}|m{1cm}|}
\hline
\multicolumn{2}{|m{2cm}|}{($++$)-start, $\theta_2=2$} &
\multicolumn{2}{m{2cm}|}{R1-start} &
\multicolumn{2}{m{2cm}|}{R2-start} &
\multicolumn{2}{m{2cm}|}{D1-start} &
\multicolumn{2}{m{2cm}|}{D2-start} &
\multicolumn{2}{m{2cm}|}{($-+$)-start} &
\multicolumn{2}{m{2cm}|}{($-++$)-start}\\\hline
$S$-$\mathcal{E}$ &
$\mathcal{E}_1$-$\mathcal{E}_2$ &
$S$-$\mathcal{E}$ &
$\mathcal{E}_1$-$\mathcal{E}_2$ &
$S$-$\mathcal{E}$ &
$\mathcal{E}_1$-$\mathcal{E}_2$ &
$S$-$\mathcal{E}$ &
$\mathcal{E}_1$-$\mathcal{E}_2$ &
$S$-$\mathcal{E}$ &
$\mathcal{E}_1$-$\mathcal{E}_2$ &
$S$-$\mathcal{E}$ &
$\mathcal{E}_1$-$\mathcal{E}_2$ &
$S$-$\mathcal{E}$ &
$\mathcal{E}_1$-$\mathcal{E}_2$\\\hline

\begin{small}
\underline{000-1}

0100

1010

11-10

1100

1101

1111
\end{small}
&
\begin{small}

-10-1-1

\underline{-11-1-1}

00-1-1

01-1-1

01-10

11-1-1

11-10
\end{small}

&
 
 &
 
 \begin{small}
-1-1-11

-11-11

1-1-1-1

1-1-11

1-11-1

\underline{1-111}

11-11
\end{small}
 &
 \begin{small}
-100-1

\underline{0100}

0101

1-10-1

1-100

1000

1101
\end{small}

 &
 \begin{small}
-1-1-1-1

-1-11-1

-1-110

-10-1-1

\underline{-1011}

-1111
\end{small}
 &
 \begin{small}
0-100

0-101

0001

0100

\underline{0101}

1-100

1000
\end{small}

 &
\begin{small}
-1-1-1-1

-1-1-10

\underline{-1-101}

-10-1-1

-11-1-1

1-1-11

\underline{11-1-1}
\end{small}
&
\begin{small}
\underline{0100}

1-10-1

1-100

1-101

100-1

1000

1101 
\end{small}
&
\begin{small}
-1-11-1

-1-110

-1-111

-10-1-1

\underline{-1011}

-11-1-1

-1111 
\end{small}
&
\begin{small}
\underline{-1-100}

0-1-1-1

0-100

0-101

0001

1-101

1001 
\end{small}
&
\begin{small}
-1-1-1-1

-1-1-10

-1-1-11

\underline{-1-101}

-10-1-1

-10-11

\underline{-1001} 
\end{small}
&
\begin{small}
\underline{000-1}

0100

1-101

11-10

1100

1101

1111 
\end{small}
&
\begin{small}
\underline{1-101}

100-1

1001

11-11

\underline{110-1}

111-1

1110 
\end{small}
\\\hline
\end{tabular}
\caption{Diversity of excitation is measured by selecting seven functions with highest values $\mathcal{E}$ and 
$S$ (columns $\mathcal{E}$-$S$) and 
diversity of excitation interval configuration by selecting seven functions with highest values $\mathcal{E}_1$ and $\mathcal{E}_2$ (columns $\mathcal{E}_1$-$\mathcal{E}_2$) , 
i.e. Shannon entropy calculated on configurations of $\theta_1$ and $\theta_2$. Functions exhibiting highest morphological diversity in their groups
are underlined.}
\end{table}
\label{diversitytable}
\end{landscape}


Top seven functions showing highest values of diversity indices, e.g. those encircled in examples 
Fig.~\ref{simpson_shanon_randomdisc_T1=1}--\ref{shannon1_shanon2_singleton_T1=2}, are 
grouped in Tab.~\ref{diversitytable} for various scenarios of initial start.

\begin{finding}
There is no match between functions which produce configurations of excitation with highest morphological 
diversity and configurations of interval boundaries with highest morphological diversity.. 
\end{finding}

\begin{figure}[!tbp]
\centering
\subfigure[($-++$)-start, $E(000-1)$]{\includegraphics[width=0.9\textwidth]{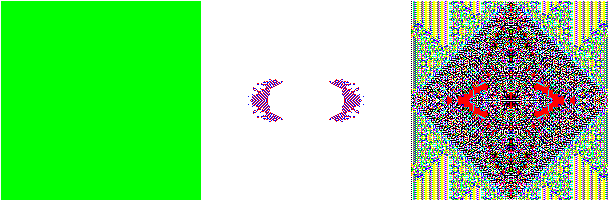}}
\subfigure[($-++$)-start, $E(1-101)$]{\includegraphics[width=0.9\textwidth]{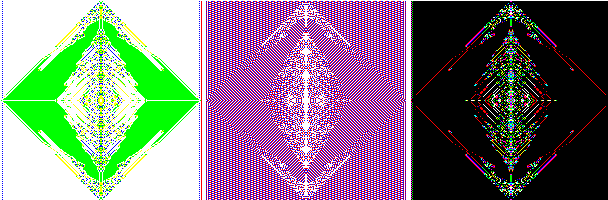}} 
\subfigure[($-++$)-start, $E(110-1)$]{\includegraphics[width=0.9\textwidth]{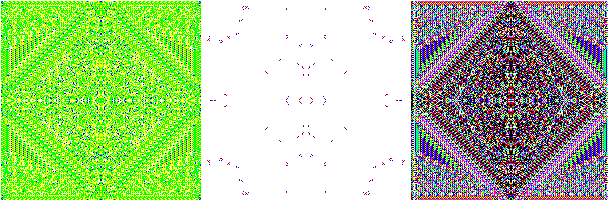}} 
\caption{Examples of most morphologically diverse configurations generated in ($-++$)-start. 
(a)~Configurations with highest morphological diversity of excitation generated by function $E(000-1)$.
(b)~Configurations with highest morphological diversity of interval boundaries $\theta_1$ and $\theta_2$ generated by function $E(1-101)$.
(c)~Configurations with highest morphological diversity of interval boundaries $\theta_1$ and $\theta_2$ generated by function $E(110-1)$.
Configurations of $\theta_1$ (left), excitation (middle) and $\theta_2$ (right) are taken in $200 \times 200$ cell array,
at $t=1000$.}
\label{examplesofmostdiverse(--+)start}
\end{figure}

Amongst functions listed in Tab.~\ref{diversitytable} only function $E(1-101)$ gets into top seven functions with highest
diversity of both excitation and interval boundaries for scenario ($-++$)-start.  Exemplar configurations of excitation and 
interval boundaries generated by automaton governed by $E(1-101)$ are shown in Fig.~\ref{examplesofmostdiverse(--+)start}b. 
Function $E(1-101)$ is also amongst functions with highest diversity for $D2$- and $(-+)$-start.   The function governs the following update of
the excitation interval boundaries.  Low boundary  $\theta_1(x)$ of excitation of cell $x$ is updated only cell $x$ is excited. The boundary
 $\theta_1(x)$ increases if cell $x$ has more excited neighbours than refractory neighbours, $\theta_1(x)$ decreases if number of refractory neighbours
 of $x$ exceeds number of excited neighbours. The boundary  $\theta_1(x)$ does not change if cell $x$ has the same number of excited neighbours as 
 refractory neighbours. Upper boundary $\theta_2(x)$ increases if cell $x$'s dissents with excitation-refractoriness ratio in its neighbourhood: 
 $x^t = +$ and $\sigma^t_+(x) < \sigma^t_-(x)$ or  $x^t = -$ and $\sigma^t_+(x) > \sigma^t_-(x)$. The boundary $\theta_2(x)$ decreases
 if cell $x$ conforms to excitation-refractoriness ratio in its neighbourhood:  $x^t = +$ and $\sigma^t_+(x) > \sigma^t_-(x)$ or  $x^t = -$ and 
 $\sigma^t_+(x) < \sigma^t_-(x)$. Increase of $\theta_1$ and decrease of $\theta_2$ lead to decrease cell's excitability. Thus we can characterise
 function $E(1-101)$ as follows: excitability of a cell decreases if the cell dissents with its neighbourhood and increases otherwise.

\begin{figure}[!tbp]
\centering
\subfigure[$++$-start, $E(000-1)$]{\includegraphics[width=0.9\textwidth]{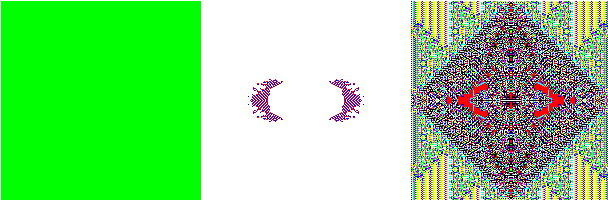}}
\subfigure[$++$-start, $E(-11-1-1)$]{\includegraphics[width=0.9\textwidth]{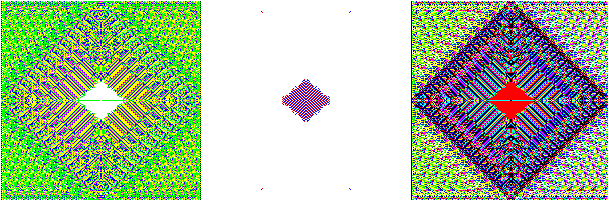}} 
\subfigure[R1-start, $E(1-111)$]{\includegraphics[width=0.9\textwidth]{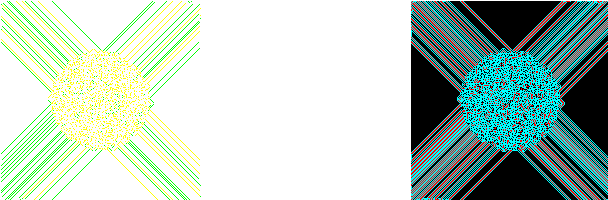}} 
\caption{Examples of most morphologically diverse configurations generated in ($++$)-start (ab) and R1-start (c). 
(a)~Configurations with highest morphological diversity of excitation generated by function $E(000-1)$.
(b)~Configurations with highest morphological diversity of interval boundaries $\theta_1$ and $\theta_2$ generated by function $E(-11-1-1)$.
(c)~ Configurations with highest morphological diversity of interval boundaries $\theta_1$ and $\theta_2$ generated by function $E(1-111)$.
 Configurations of $\theta_1$ (left), excitation (middle) and $\theta_2$ (right) are taken in $200 \times 200$ cell array,
at $t=1000$.}
\label{examplesofmostdiverse1}
\end{figure}

\begin{figure}[!tbp]
\centering
\subfigure[R2-start, $E(0100)$]{\includegraphics[width=0.9\textwidth]{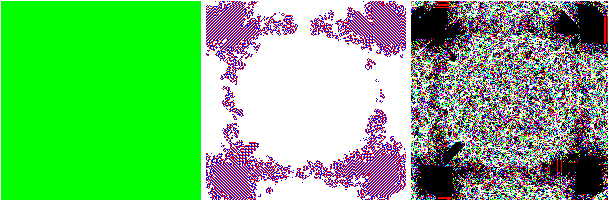}} 
\subfigure[R2-start, $E(-1011)$]{\includegraphics[width=0.9\textwidth]{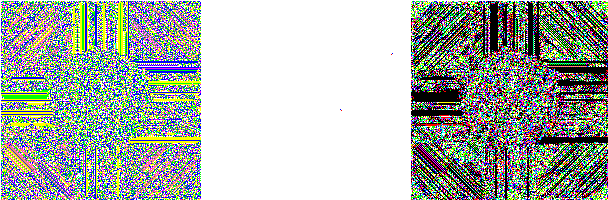}} 
\caption{Examples of most morphologically diverse configurations generated in R2-start. 
(a)~Configurations with highest morphological diversity of excitation generated by function $E(0100)$.
(b)~Configurations with highest morphological diversity of interval boundaries $\theta_1$ and $\theta_2$ generated by function $E(-1011)$.
Configurations of $\theta_1$ (left), excitation (middle) and $\theta_2$ (right) are taken in $200 \times 200$ cell array,
at $t=1000$.}
\label{examplesofmostdiverseR2start}
\end{figure}

\begin{figure}[!tbp]
\centering
\subfigure[D1-start, $E(0101)$]{\includegraphics[width=0.9\textwidth]{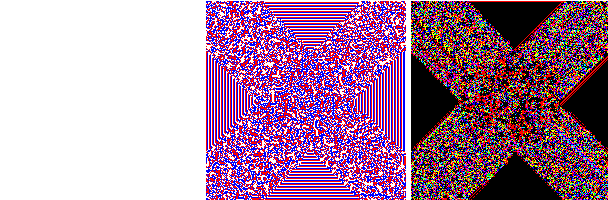}} 
\subfigure[D1-start, $E(-1-101)$]{\includegraphics[width=0.9\textwidth]{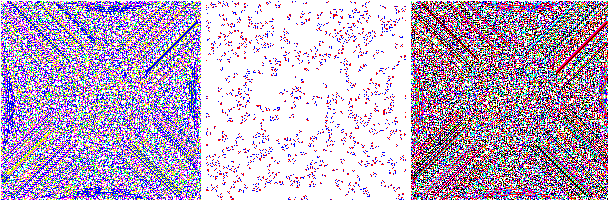}} 
\subfigure[D1-start, $E(11-1-1)$]{\includegraphics[width=0.9\textwidth]{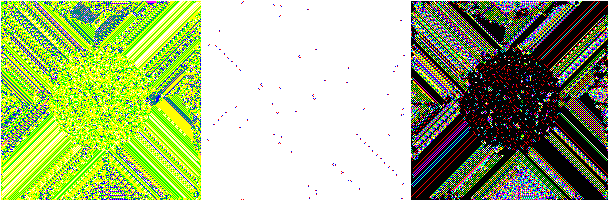}} 
\caption{Examples of most morphologically diverse configurations generated in D1-start. 
(a)~Configurations with highest morphological diversity of excitation generated by function $E(0101)$.
(b)~Configurations with highest morphological diversity of interval boundaries $\theta_1$ and $\theta_2$ generated by function $E(-1-101)$.
(c)~Configurations with highest morphological diversity of interval boundaries $\theta_1$ and $\theta_2$ generated by function $E(11-1-1)$.
Configurations of $\theta_1$ (left), excitation (middle) and $\theta_2$ (right) are taken in $200 \times 200$ cell array,
at $t=1000$.}
\label{examplesofmostdiverseD1start}
\end{figure}

\begin{figure}[!tbp]
\centering
\subfigure[D2-start, $E(0100)$]{\includegraphics[width=0.9\textwidth]{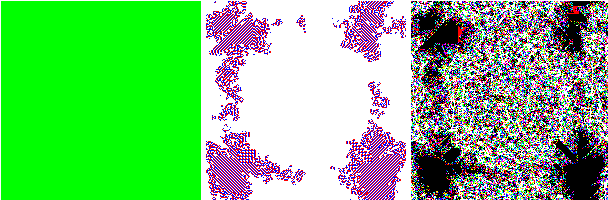}} 
\subfigure[D2-start, $E(-1011)$]{\includegraphics[width=0.9\textwidth]{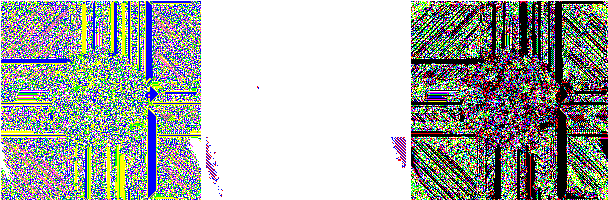}} 
\caption{Examples of most morphologically diverse configurations generated in D2-start. 
(a)~Configurations with highest morphological diversity of excitation generated by function $E(0100)$.
(b)~Configurations with highest morphological diversity of interval boundaries $\theta_1$ and $\theta_2$ generated by function $E(-1011)$.
Configurations of $\theta_1$ (left), excitation (middle) and $\theta_2$ (right) are taken in $200 \times 200$ cell array,
at $t=1000$. }
\label{examplesofmostdiverseD2start}
\end{figure}

\begin{figure}[!tbp]
\centering
\subfigure[($-+$)-start, $E(-1-100)$]{\includegraphics[width=0.9\textwidth]{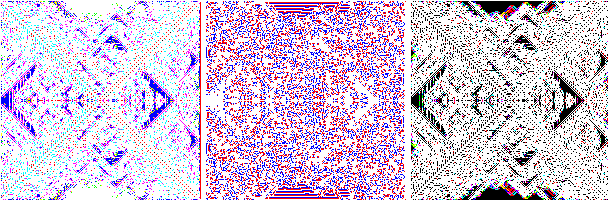}}
\subfigure[($-+$)-start, $E(-1-101)$]{\includegraphics[width=0.9\textwidth]{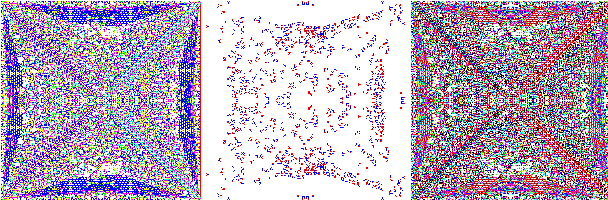}} 
\subfigure[($-+$)-start, $E(-1001)$]{\includegraphics[width=0.9\textwidth]{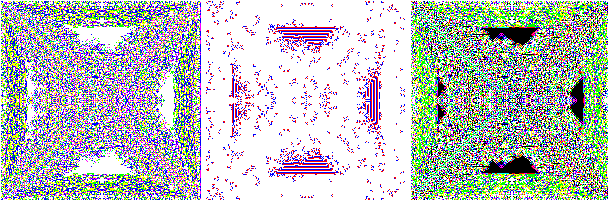}} 
\caption{Examples of most morphologically diverse configurations generated in ($-+$)-start. 
(a)~Configurations with highest morphological diversity of excitation generated by function $E(-1-100)$.
(b)~Configurations with highest morphological diversity of interval boundaries $\theta_1$ and $\theta_2$ generated by function $E(-1-101)$.
(c)~Configurations with highest morphological diversity of interval boundaries $\theta_1$ and $\theta_2$ generated by function $E(-1001)$.
Configurations of $\theta_1$ (left), excitation (middle) and $\theta_2$ (right) are taken in $200 \times 200$ cell array,
at $t=1000$.}
\label{examplesofmostdiverse(-+)start}
\end{figure}

\begin{finding}
Functions $E(-1-101)$ and $E(-1011)$ generate  configurations of excitation interval boundaries 
with highest morphological diversity for three and two types of initial stimulation, respectively, 
and functions $E(000-1)$ and $E(0100)$  generate configurations of excitation with
highest morphological diversity for two types of initial stimulation.
\end{finding}

The function $E(-1-101)$ generates configurations with highest morphological diversity of 
$\theta_1$ and $\theta_2$ configurations for   R2-, D1- and ($-+$)-starts, see examples
in Fig.~\ref{examplesofmostdiverseD1start}b and \ref{examplesofmostdiverse(-+)start}b, and 
$E(-1011)$ generates highest morphological diversity  configurations of $\theta_1$ and $\theta_2$
for R2- and D2-starts (Figs.~\ref{examplesofmostdiverseR2start}b and \ref{examplesofmostdiverseD2start}b). 
The functions  $E(000-1)$ and $E(0100)$ produce highest morphological diversity configurations of excitation 
for  ($++$)- and ($--+$)-starts ($E(000-1)$) and R2- and D2-starts ($E(0100)$). See examples for  $E(000-1)$ in
Figs.~\ref{examplesofmostdiverse1}a and \ref{examplesofmostdiverse(--+)start}a and  $E(0100)$ in 
Figs.~\ref{examplesofmostdiverseR2start}a and \ref{examplesofmostdiverseD2start}a.

\begin{finding}
Function $E(0100)$ generates most morphologically diverse excitation patterns in larger, comparing to other functions, 
number of initial conditions.
\end{finding}

Function $E(0100)$ is amongst top seven functions with highest morphological diversity of excitation in 
($++$)-,  R2-,  D1-, D2- and ($-++$)-starts (Tab.~\ref{diversitytable}). The function generates most 
morphologically diverse excitations in D2-start. Examples of configurations generated by $E(0100)$ 
are shown in Figs.~\ref{examplesofmostdiverseR2start}a and \ref{examplesofmostdiverseD2start}a.
The function $E(0100)$ shows how dynamics of excitation can be tuned by changing only 
upper boundary of the excitation interval with lower boundary fixed. Value $\theta_1(x)$ is not updated. 
Value $\theta_2(x)$ increases if cell $x$ is excited and it has more excited than refractory neighbours, 
the value $\theta_2(x)$ decreases if cell $x$ is excited and has more refractory than excited neighbours. 
Excitability of a cell decreases if the cell dissents with excitation ratio in its neighbourhood, and 
increases otherwise.

\begin{finding}
Function $E(-10-1-1)$ generates  most morphologically diverse patterns of interval boundaries  in larger, 
comparing to other functions, number of initial conditions.
\end{finding}

Function $E(-10-1-1)$ gets in top seven functions with highest morphological diversity of $\theta_1$ and $\theta_2$ patterns in 
($++$)-, R2-, D1-, D2-, and ($-+$)-starts (Tab.~\ref{diversitytable}). In automata, governed by this function, 
$\theta_2(x)$ is not updated if cell $x$ is excited. Otherwise, $\theta_1(x)$ and $\theta_2(x)$ increase if 
refractory neighbours outnumber in the cell $x$'s neighbourhood and decrease if excited neighbours outnumber refractory neighbours.

\begin{finding}
Upper boundary of excitation interval more significantly affects morphological diversity of configurations generated 
than lower boundary of the interval does.
\end{finding}

There are only two functions, $E(1010)$ and $E(1000)$,  where only lower boundary $\theta_1$ is updated in Tab.~\ref{diversitytable}.
Function $E(1010)$ represents a situation when $\theta_1(x)$ is independently of a state of cell $x$: $\theta_1(x)$ increases if number 
of excited neighbours exceeds number of refractory neighbours, and $\theta_1(x)$ decreases if refractory neighbours outnumber excited
neighbours.  In automata
governed by function $E(1000)$ value of $\theta_1(x)$ is updated as above but only if  cell $x$ is excited. There are several 
functions with highest morphological diversity which represent fixed lower boundary and dynamical upper boundary, e.g. 
$E(000-1)$ (Figs.~\ref{examplesofmostdiverse1}a and \ref{examplesofmostdiverse(--+)start}a), 
$E(0100)$ (Figs.~\ref{examplesofmostdiverseR2start}a and \ref{examplesofmostdiverseD2start}a), 
$E(0101)$ (Fig.~\ref{examplesofmostdiverseD1start}a), 
$E(0-100)$, $E(0-101)$, and $E(0001)$ (Tab.~\ref{diversitytable}).

\section{Generative diversity and localisations}
\label{Generativediversityandlocalisations}

\begin{figure}[!tbp]
\centering
\includegraphics[width=\textwidth]{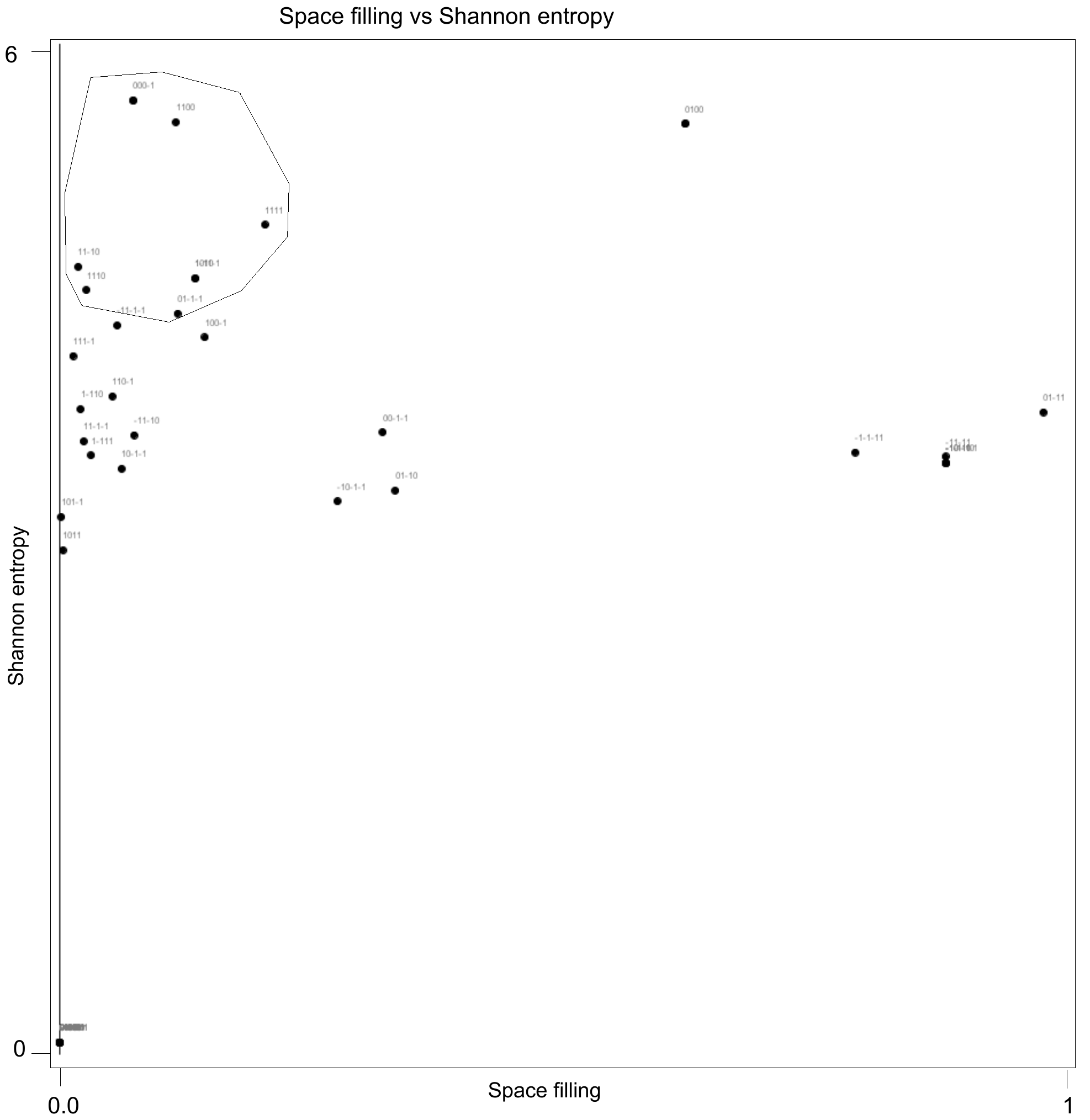}
\caption{Generative diversity of functions for ($++$)-start, $\theta^0_1(x)=2$ for all $x$: 
Space filling (horizontal axis) vs Shannon entropy (vertical axis) for configuration of excitable array 
of $200 \times 200$ cells with periodic boundary condition, recorded at $t=1000$. 
Encircled data points are seven functions with highest generative diversity specified in column ($++$)-start, 
in Tab.~\ref{generativediversity}.}
\label{spaceFillingvsShannonSingletonT1=2}
\end{figure}

\begin{figure}[!tbp]
\centering
\subfigure[($++$)-start, $\theta^0_1=2$, $E(1111)$]{\includegraphics[width=0.9\textwidth]{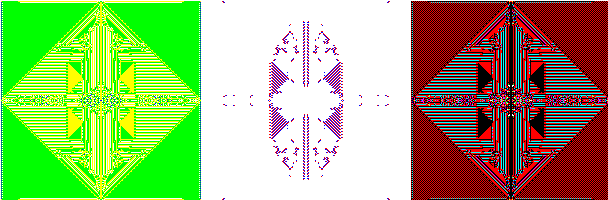}}
\subfigure[($++$)-start, $\theta^0_1=2$, $E(11-10)$]{\includegraphics[width=0.9\textwidth]{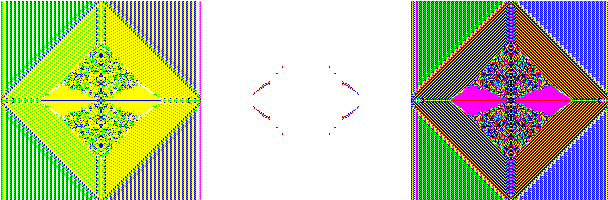}} 
\subfigure[($-+$)-start, $\theta^0_1=1$, $E(0-1-1-1)$]{\includegraphics[width=0.9\textwidth]{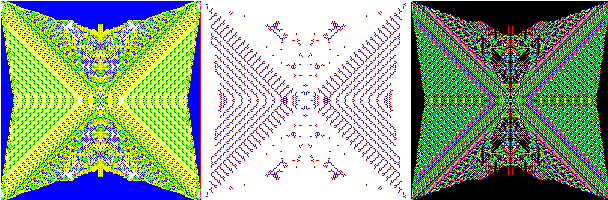}} 
\caption{Examples of configurations generated by functions with highest generative diversity.
(a)~$E(1111)$, (b)~$E(11-10)$, (c)~$E(0-1-1-1)$. Automaton array has $200 \times 200$ cells, 
configurations of $\theta_1$ (left), excitation (middle) and $\theta_2$ (right) at $t=1000$.}
\label{examples(1111)(11-10)(0-1-1-1)}
\end{figure}

Generative diversity characterises how many different configurations are generated during space-time development
of automaton starting with entirely resting configuration but single cell in a non-resting state. We consider two starting 
conditions: ($++$)-start and $\theta_1^0(x)=2$ for any $x$;  ($-+$)-start and $\theta_1^0(x)=1$; and,
($--+$)-start and  $\theta_1^0(x)=2$. Generative diversity is evaluated using Shannon entropy and space-filling (a ratio of 
cells in a non-resting state). Functions generating configurations with maximum Shannon entropy and minimum 
space-filling are assumed to have higher generative complexity, see example in  Fig.~\ref{examples(1111)(11-10)(0-1-1-1)}.

\begin{table}[!tbp]
\begin{tabular}{c|c}
($++$)- and  ($-++$)-start, $\theta^0_1=2$ & ($-+$)-start, $\theta^0_1=1$ \\ \hline
\begin{math}
\begin{array}{l}
-11-1-1\cr
01-1-1\cr
100-1\cr
1010\cr 
11-10\cr
1110\cr
1111
\end{array}
\end{math}
&
\begin{math}
\begin{array}{l}
-1-101 \cr
0-1-1-1\cr
00-1-1\cr
01-1-1\cr
01-11\cr
10-1-1\cr
11-11
\end{array}
\end{math}
\end{tabular}
\caption{Seven functions with highest generative diversity for ($++$)- and  ($-++$)-start, $\theta_1=2$ (first column) 
and ($-+$)-start, $\theta_1=1$ (second column).}
\label{generativediversity}
\end{table}

Seven functions with highest generative diversity are listed in Tab.~\ref{generativediversity}.
Configurations of excitation and interval boundaries for 
$E(-11-1-1)$, $++$-start, shown in Fig.~\ref{examplesofmostdiverse1}b,
$E(-1-101)$, ($-+$)-start, in Fig.~\ref{examplesofmostdiverse(-+)start}b and 
configurations generated by functions $E(1111)$, $E(11-10)$ and $E(0-1-1-1)$ in 
Fig.~\ref{examples(1111)(11-10)(0-1-1-1)}.

\begin{table}[!tbp]
\begin{tabular}{cc}
($++$)-start & ($-+$)-start \\ \hline
 \begin{math}
\begin{array}{l}
10-1-1 \cr
11-1-1 \cr
11-10 \cr
100-1 \cr
101-1 \cr
110-1 \cr
111-1 \cr
1011 \cr
1110 
 \end{array}
\end{math}
&
\begin{math}
\begin{array}{l}
-10-1-1 \cr
-11-10 \cr
00-1-1 \cr
01-1-1 \cr
01-10 \cr
10-1-1 \cr
11-1-1 \cr
11-10
 \end{array}
\end{math}
\end{tabular}
\caption{Functions supporting localizations in ($++$)- and ($-++$)-start (left column) and ($-+$)-start (right column).}
\label{functionslocalisations}
\end{table}

\begin{finding}
Most localizations generated in  ($++$)- and ($-++$)-starts are stationary.
\end{finding}

\begin{figure}[!tbp]
\centering
\subfigure[($++$)-start, $\theta^0_1=2$, $E(00-1-1)$]{\includegraphics[width=0.9\textwidth]{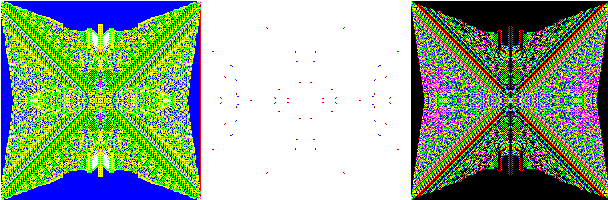}}
\subfigure[($++$)-start, $\theta^0_1=2$, $E(101-1)$]{\includegraphics[width=0.9\textwidth]{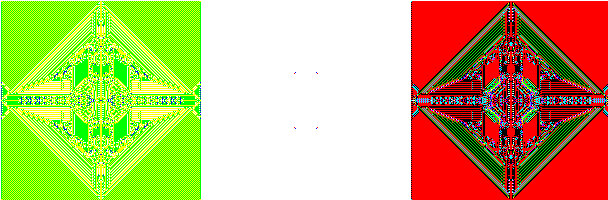}}
\subfigure[($++$)-start, $\theta^0_1=2$, $E(1110)$]{\includegraphics[width=0.9\textwidth]{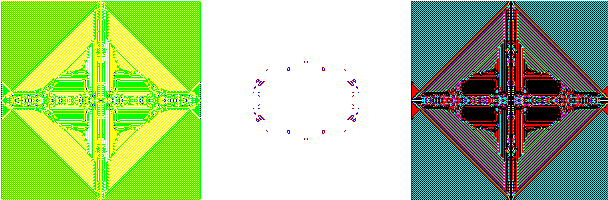}}
\subfigure[($-+$)-start, $\theta^0_1=1$, $E(00-1-1)$]{\includegraphics[width=0.9\textwidth]{figs/localisations/00m1m1_36}}
\caption{Examples of configurations with localised excitations developed in ($++$)-start (abc) and ($-+$)-start (d) scenarios. 
Size of cellular array is $200 \times 200$ cells, configurations of $\theta_1$ (left), excitation (middle) and $\theta_2$ (right)
at $t=1000$.}
\label{exampleslocalisations}
\end{figure}

Around half of the  functions  generate configurations with localizations in case of R2-start (43), D1-start (41), 
D2-start (43). We concentrate on  functions which produce localizations in singleton starts. There nine functions for 
($++$)- and eight function for  ($-+$)-start , $\theta^0_1=2$ (Tab.~\ref{functionslocalisations}). Examples of configurations
generated by functions $E(00-1-1)$,  $E(101-1)$, $E(1110)$ and $E(00-1-1)$ are shown in Fig.~\ref{exampleslocalisations}.

\begin{finding}
Functions $E(11-10)$ and $100-1$ are amongst top seven functions with highest generative diversity supporting localised excitation dynamics
in scenarios of ($++$)-start. Functions $E(00-1-1)$, $E(01-1-1)$, $E(10-1-1)$ and $E(11-1-1)$  are amongst top seven functions with highest 
generative diversity supporting localised excitation dynamics in scenarios of ($-+$)-start
\end{finding}

See Tabs.~\ref{generativediversity} and \ref{functionslocalisations}. Configurations generated by function $E(11-10)$ 
are exemplified in Fig.~\ref{examples(1111)(11-10)(0-1-1-1)}b and function $E(00-1-1)$ in 
Fig.~\ref{exampleslocalisations}a. Functions $E(100-1)$ and $E(00-1-1)$ are the functions with minimal updates of 
excitation interval boundaries. In automata governed by function $E(00-1-1)$ $\theta_1(x)$ and $\theta_2(x)$ are updated
only if the cell $x$ is in refractory state: both boundaries decrease if excited neighbours of $x$ outnumber refractory neighbours, 
and they increase refractory neighbours dominate. In automata governed by function $E(100-1)$ the boundary $\theta_1(x)$ 
increases if the excited cell $x$ has more excited neighbours than refractory ones, and boundary and decreases if the cell
has more refractory neighbours. The boundary $\theta_2(x)$ is updated only if cell $x$ is refractory: $\theta_2(x)$ decreases if 
excited neighbours outnumber refractory neighbours, and it increases otherwise.

\section{Summary}
\label{summary}

Excitable cellular automata with dynamical excitation interval exhibit a wide range of space-time dynamics based 
on an interplay between  propagating excitation patterns which modify excitability of the automaton cells. Such 
interactions leads to formation of standing domains of excitation, stationary waves and localised excitations. We analysed 
morphological and generative diversities of the functions studied and characterised the functions with highest values of 
the diversities. Amongst other intriguing discoveries we found that upper boundary of excitation interval more significantly affects morphological diversity of  configurations generated than lower boundary of the interval does and there is no  match between functions which produce configurations of excitation with highest morphological diversity and configurations of interval boundaries with highest morphological diversity.  Potential directions of futures studies of excitable media with dynamically changing excitability may focus  on relations of the automaton model with living excitable media, e.g. neural tissue and muscles, and novel materials with memristive properties, and networks of conductive polymers.

\end{document}